\documentclass[11pt]{cernrep}
\usepackage{graphicx}
\usepackage{cite,./mcite}
\usepackage{wrapfig,rotating}
\usepackage{epsf}
\usepackage{epsfig}
\usepackage{cite}
\usepackage{url}

\begin{document}
\title{HEP data analysis using jHepWork and Java}

\author{S.~Chekanov}
\documentlabel{ANL-HEP-CP-08-53}

\institute{
HEP Division, Argonne National Laboratory,
9700 S.Cass Avenue,
Argonne, IL 60439
USA \\
}

\maketitle
\begin{abstract}
A role of Java in high-energy physics (HEP) and recent progress
in development of a  platform-independent data-analysis framework, jHepWork, is discussed.
The framework produces professional graphics and has many 
libraries for data manipulation.
\end{abstract}

\section{Introduction}

Nowadays, the advantages of Java over C++ seem overwhelming.
Being the most popular open-source programing 
language\footnote{According to SourceForge.net and Freshmeat.net statistics, 
the number of open-source applications written in Java exceeds those written in C++.},  
Java retains the C++ syntax, but significantly simplifies the language.
This is (incomplete) list of advantages of Java over C++:
1) Java is multiplatform with the philosophy of “write once, run anywhere";  
2) Better structured, clean, efficient, simpler (no pointers);
3) Stable, robust and well supported:  
Java programs written (or compiled)
many years from now can be compiled (or executed) without modifications even today.
This is true even for JAVA source code with graphic widgets.  
In contrast, C++ programs 
always require continues time-consuming maintenance in order to follow 
the development of C++ compilers and graphic desktop
environment; 
4) Java has reflection technology, which is not present in C++. The reflection
allows an application to discover information about created objects, 
thus a program can  design  
itself at runtime. In particular, this is considered to be essential
for building "intelligent" programs making decisions at runtime; 
5) Free intelligent IDEs, which are absolutely
necessary for large software projects\footnote{For example, 
the total number of lines of source code in ATLAS software is far higher than hundreds of
thousands lines.};
6) Automatic garbage collection, i.e.  a programmer does not need to perform memory management; 
7) Extensive compile-time and run-time checking; 8) Programs written in Java
can be embedded to the Web. This is  
important for distributed analysis environment (Java webstart, plugins, applets),
especially when HEP data analysis tools are not localized in one single laboratory but
scattered over the Web.

The importance of Java in HEP 
data analysis has been recognized since establishing the FreeHEP
Java library and producing a first version of JAS (Java analysis studio)~\cite{jas}. 
Presently, many elements of the grid software are written
in Java. At LHC, Java is used for event displays and several other areas. 
While C++ language is remaining to be the main programming language at LHC,
it lacks many features existing in Java, which makes the entire LHC software environment
tremendously complicated and difficult to handle by the end users. 
The lack of robustness and backward compatibility of
C++ free compilers leads
to various HEP-supported "scientific" flavors of Linux, 
with different architecture (32 bit or 64), which are all 
tightened to particular libraries and hardware.
For example,  the main computational platform for ATLAS is Scientific Linux 4.6. It will be
used for future data taking, however,  even now it 
is several generations behind the main-stream Linux 
modern distributions (Fedora, Ubuntu, Suse  etc) and 
cannot be easily installed on modern laptops. 
Currently, the HEP community is required to support the entire computing chain, 
from hardware and operating systems,
to the end-user programs,  rather than concentrating on HEP-specific computational tasks.
This is a significant difference from the initial concept, when HEP software 
could be run essentially on any platform and a vendor-supported operating system.

It should be pointed out that C+ 
has been chosen as the main programming language at LHC at the time when Java
was still behind C++, lacking Just-in-time (JIT) compilers to convert parts of the bytecode 
to native code in order to improve execution time. 
At that time, Python~\cite{python_web}, another portable programming language,
also did not have enough  power to be widely used in HEP.
As Java, Python has also become increasingly
popular programming language in science
and engineering~\cite{python}, since it is interactive, object-oriented,
high-level, dynamic and
portable. It has simple and easy to learn syntax
which reduces the cost of program maintenance.
While being portable, Python implemented in C (CPython) requires
user-specific C/C++ libraries for high-performance computing, thus it cannot 
be considered a basis for a multiplatform data-analysis environment.

Jython~\cite{jython_web} is an implementation of Python in Java
and, as any Java application, is  truly multiplatform.
In contrast to CPython,
Jython is fully integrated with the Java platform, thus
Jython programs can make full use of
extensive built-in and third-party Java libraries.
Therefore, Jython programs have even more power than
the standard Python implemented in C.
Finally, the Jython interpreter
is freely available for both commercial and non-commercial use.

jHepWork~\cite{jhepwork} is a full-featured object-oriented data analysis
framework for scientists that takes advantage of the
Jython language and Java.
Jython macros are used for data manipulation, data  visualization (plotting 
1D and 2D histograms), statistical analysis, fits, etc.
Data structures and data manipulation methods
integrated with Java and JAIDA FreeHEP libraries~\cite{freehep}
combine remarkable power with a very clear syntax.
jHepWork Java libraries can also be used to develop programs using the standard JAVA,
without Jython macros.

Programs written using jHepWork are usually rather short due the
simple Python syntax and high-level
constructs implemented in
the core jHepWork libraries.
As a front-end data-analysis environment,
jHepWork helps to concentrate on interactive
experimentation, debugging, rapid script development and finally on
workflow of scientific tasks, rather than
on low-level programming.

jHepWork is an open source product which is implemented 100 percent in Java.
Since it is fully multiplatform, it does
not require installation and can be run on any platform where Java is installed.
It can be used to develop a range of data-analysis applications focusing
on analysis of complicated data sets, histograms, statistical analysis of data,
fitting. It offers a full-featured, extensible
multiplatform IDE implemented in Java.

jHepWork is seamlessly integrated with Java-based Linear Collider Detector (LCD)
software concept and it has the core based using FreeHEP libraries and other GNU-licensed
packages.  
While jHepWork is mainly designed to be used in high-energy physics, it can also be
used in any field, since all methods and classes are rather common
in science and engineering.

Below we will discuss only the key features of jHepWork,
without the coverage of all available methods, which can easily be found using an extensive
help system and the code completion feature of jHepWork.
The main web page of jHepWork~\cite{jhepwork} 
contains the package itself, user manuals and about 50 examples with various macros.
jHepWork consists of two major libraries: jeHEP (jHepWork IDE) and jHPlot
(jHepWork data-analysis library). Both are
licensed by the GNU General Public License (GPL).

\section{Main differences with other data-analysis tools}

Below we will compare jHepWork with two popular object-oriented packages currently
used in high-energy physics:
1) JAS package \cite{jas}, based on Java and FreeHEP libraries \cite{freehep} and
2) C++ ROOT package~\cite{Brun:2003ga,*Brun:1997pa}.

\subsection{Main differences with JAS}

Compare to JAS, jHepWork:

\begin{itemize}

\item
has a full-featured integrated development environment (IDE) with
syntax highlighting,
syntax checker, code completion, code analyser, an Jython shell and a file manager. 

\item
contains powerful libraries to display data (including 3D plots) with a large
choice for interactive labels and text attributes (subscripts, superscripts, overlines,
arrows, Greek symbols etc.).
jHepWork plots are more  interactive than those written using 
FreeHEP JAIDA libraries linked with JAS.
The plotting part is based on the jHPlot
library developed for the jHepWork project and JaxoDraw Java application \cite{jaxodraw}.
The latter can be used  to draw Feynman diagrams in addition to standard plots;

\item
is designed to write short programs due to several enhancements
and simpler class names. The classes written for jHepWork were
designed keeping in mind simplicity of
numerous high-level constructs enabling the user
to write programs that are significantly shorter than programs written using JAS;

\item
includes high-level constructions for data manipulations, data presentations
in form of tables, data input and output,  calculations
of systematical errors and visualization (plots, tables, spreadsheet, neural networks)
which have no analogy in JAS;

\item
includes an advanced help system with the code completion.
For the core jHplot package, the  code completion feature is complimented with a detailed
API information on each method associated with certain class.

\end{itemize}

\subsection{Main differences with the ROOT  package}

Compare to ROOT, jHepWork:

\begin{itemize}

\item
is seamlessly integrated with Java-based Linear Collider Detector (LCD)
software concept;

\item
is a Java-based program, thus it is fully multiplatform and does
not require installation.
This is especially useful for plugins distributed via the Internet in
form of bytecode jar libraries;

\item
Java is very robust.
Java source codes developed many years from now  can easily be compiled without any
changes even today. 
Even class libraries compiled many years from now can run on
modern Java Virtual Machines. Therefore, the maintenance of jHepWork 
package is
much less serious issue compared to ROOT;

\item
since jHepWork is 100\% Java, it has automatic garbage collection, which is
significant advantage over ROOT C++/C;

\item
has a full-featured IDE with
syntax highlighting,
syntax checker, code completion and analyser;

\item
can be integrated with the Web in form of applets, thus
it is better suited for distributed analysis
environment via the Internet. This is essential
feature for modern large collaborations in high-energy physics and in other
scientific fields;

\item
calculations based on Jython/Python scripts are typically 4-5
times shorter than equivalent ROOT/C++ programs. 
Several examples are discussed in Ref.~\cite{jhepwork};

\item
calculations based on Jython scripts can be compiled to Java
bytecode files and packed to jar class libraries without
modifications of Jython scripts.  In contrast,  ROOT/CINT scripts
have to be written using a proper C++ syntax, without CINT shortcuts,
if they will be compiled into
shared libraries; 

\item
can access high-level Python and Java data structures;

\item
includes an advanced help system with a code completion based
on the Java reflection technology.  With increasingly large number of
classes and methods in ROOT,
it is difficult to understand which method belongs to which particular class.
Using the jHepWork IDE, it is possible to access the full
description of all classes and methods
during editing Jython scripts;

\item
automatic updates which does not depend on particular platform.
For ROOT, every new version has to be compiled from scratch;

\item
powerful and intelligent external IDEs (Eclipse, NetBean etc) can be used 
productivity in developing HEP analysis.

\end{itemize}

\subsection{How fast it is?}

Jython scripts are about 4-8 times slower than equivalent Java programs and
about a factor five slower than the equivalent ROOT/CINT codes for operations
on primitive data types (remember, all Jython data types are objects).
This means that CPU extensive tasks should be moved to Java jar libraries.

jHepWork was designed for a data analysis in  which program
speed is not essential, as it is assumed that JHepWork scripts are used for
operations with data and objects (like histograms)  which have alredy been
created by C++, Fortran or Java code.
For such front-end data analysis, the bottleneck is mainly
user input speed, interaction with a graphical object using mouse or network latency.

In practice, final results obtained with Jython programs can be obtained
much faster than those designed in C++/Java,
because development is so much easier in jHepWork that a user often winds up with a
much better algorithm based on Jython syntax and jHepWork high-level
objects than he/she would in C++ or Java.
In case of CPU extensive tasks, like large loops over primitive data types,
reading files etc. one should use high-level structures of Jython and jHepWork or
user-specific libraries which can be developed using the jHepWork IDE.
Many examples are discussed in the jHepWork manual~\cite{jhepwork}. 

Acknowledgments. I would like to thanks many people for support,  ideas  and debugging 
of the current jHepWork version. 
This work supported in part by the U.S. Department of Energy, Division of High Energy Physics, under Contract DE-AC02-06CH11357. 

\bibliographystyle{elsart-num}
{\raggedright
\bibliography{jhepwork}
}
\end{document}